\newcommand{\be}{\begin{eqnarray}}
\newcommand{\ee}{\end{eqnarray}}
\newcommand{\nn}{\nonumber}

\documentclass[12pt]{article}

\usepackage{amsmath,amssymb}
\usepackage{graphicx}
\usepackage{cite}
\usepackage{bm}
\begin{document}

\renewcommand{\thefootnote}{\fnsymbol{footnote}}

\vskip 15mm

\begin{center}

{\Large Physics of crypto-Hermitian and/or cryptosupersymmetric field theories}

\vskip 4ex

A.V. \textsc{Smilga}\,$^{1}$,

\vskip 3ex

$^{1}\,$\textit{SUBATECH, Universit\'e de
Nantes,  4 rue Alfred Kastler, BP 20722, Nantes  44307, France
\footnote{On leave of absence from ITEP, Moscow, Russia.}}
\\
\texttt{smilga@subatech.in2p3.fr}
\end{center}

\vskip 5ex

\begin{abstract}
\noindent  We discuss non-Hermitian field theories where the spectrum of the Hamiltonian 
involves only real energies. We
make three observations. {\it (i)} The theories obtained from supersymmetric theories by 
non-anticommutative  deformations belong
in many cases to this class. {\it (ii)} When the deformation parameter is small, the deformed 
theory enjoys {\it the same} supersymmetry
algebra as the undeformed one. Half of the supersymmetries are manifest and the existence of 
another half can be deduced from the structure
of the spectrum. {\it (iii)} Generically, the conventionally defined $S$--matrix is not 
unitary for such theories.    
\end{abstract}

\setcounter{footnote}0
\setcounter{page}{1}

\section{Introduction}
 There exists a rich class of quantum systems whose Hamiltonian is apparently not Hermitian,
 but which involve
only real energies in the spectrum. Such systems have been intermittently discussed in the literature
since mid-seventies \cite{BC}. They attracted great interest after the beautiful paper
\cite{BB}, where it was shown that the Hamiltonians with a certain type of complex potentials, like
 \be
\label{HBB}  
H = p^2 + ix^3 \, , 
 \ee
have  real discrete spectrum. 

It was observed in \cite{Ali1} that all such Hamiltonians are "crypto-Hermitian", i.e. 
can be represented in the form
 \be
\label{HRH}
H \ =\ e^R \tilde{H} e^{-R}
 \ee
with Hermitian $\tilde{H}$. The matrix $e^R$ is generically not unitary and hence $H$ is not Hermitian,
but the spectrum of $H$ is the same as for $\tilde{H}$ and involves only real eigenvalues.
The similarity transformation (\ref{HRH}) amounts to redefining the metric in Hilbert space.
 The Hamiltonian $H$
is not Hermitian with respect to the standard metric, but it is Hermitian with respect to a redefined
 one. 

There is, however, a price that one has to pay for this. First, the transformation matrix $R$ 
is typically not local. For the Hamiltonian (\ref{HBB}), it involves inverse powers of momentum.
 For the Hamiltonian $H = p^2 + x^2 + igx^3$, the momentum does not show up in the denominator, but the
matrix $R$ represents an infinite series in $g$ with growing powers of momentum.
Second, the observables $x,p$, which are Hermitian with respect to the standard metric,
 are not Hermitian
with respect to the new one and could hardly be interpreted as the coordinate and momentum 
of a physical particle.

For the systems like (\ref{HBB}) with discrete real spectrum, the latter circumstance 
might be not so relevant. The spectral problem for the Hamiltonian (\ref{HBB}) {\it can} be solved
in a certain complex region of $x$. The solution is nontrivial, interesting, and one 
can be satisfied with this not asking  the question of what is the physical interpretation of $x$.
The situation is different, however, for cryptoreal systems with continuous spectrum, for example,
for the PT--symmetric square well potential
 \be
\label{well}
V(x) \ =\ \left[ 
\begin{array}{c} 
0, \ |x|>a \\ iC, \ 0< x \leq a \\  -iC, \ -a < x \leq 0 
 \end{array} \right.
  \ee
As was discussed in \cite{Ali2,Jones}, in spite of the fact that such Hamiltonian is cryptoreal
(its spectrum is real and the metric with respect to which the Hamiltonian is Hermitian exists),
the scattering matrix {\it is} not unitary if defined in a usual way as  the transition amplitude
 between ingoing and  outgoing plane waves. Cryptoreality dictates that a basis where the evolution 
operator
 is unitary exists, but the states of this basis do not have a simple physical interpretation.
This difficulty is an evident manifestation of the fact that the Hamiltonian $H$ and the observable
$x$ cannot be made Hermitian simultaneously.

\begin{figure}[h]
   \begin{center}
 \includegraphics[width=4.0in]{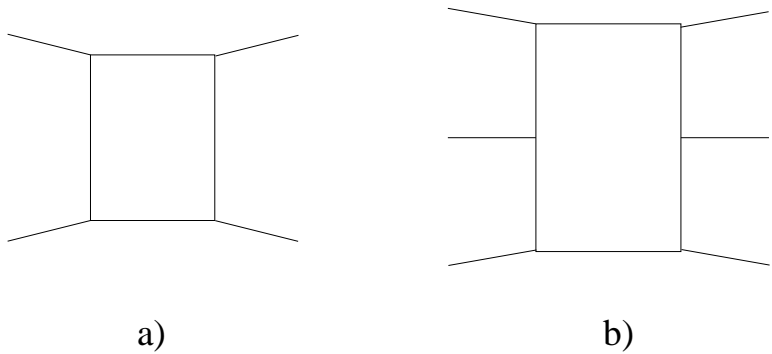}
    \end{center}
\caption{Forward scattering amplitudes in $i\phi^3$ theory.}
\label{vpered}
\end{figure}

We would like to make an obvious remark that the same applies to PT--symmetric and other cryptoreal
field theories. The simplest example of the latter is the theory of a real scalar field $\phi(x)$ with
the Lagrangian \cite{Brody}
 \be
\label{iphi3}
 {\cal L} = \frac 12 (\partial_\mu \phi)^2 - \frac 12 m^2 \phi^2 + i\gamma \phi^3\ .
\ee
The corresponding Hamiltonian is Hermitian with respect to the metric involving the transformation
matrix 
 \be
\label{Riphi3}
R \ =\ \gamma\int d{\bf x} d {\bf y} d{\bf z} \left(
M_{{\bf x}  {\bf y}{\bf z}} \pi_{\bf x} \pi_{\bf y} \pi_{\bf z} +
N_{{\bf x}  {\bf y}{\bf z}} \pi_{\bf x} \phi_{\bf y} \phi_{\bf z} \right)  + O(\gamma^3) 
 \ee
with complicated nonlocal kernels $M_{{\bf x}  {\bf y}{\bf z}}, 
N_{{\bf x}  {\bf y}{\bf z}}$ ($\pi_{\bf x}$ are the canonical momenta). That means that, when the theory
(\ref{iphi3}) is regulated in the infrared and is put in a finite spatial box, the spectrum
of the corresponding Hamiltonian is discrete and real. On the other hand, the most relevant question
 we usually ask about a quantum field system is not what is its spectrum in  finite box, but rather
what is its $S$ matrix - a set of transition amplitudes between standard {\it in} and {\it out}
states, the eigenstates of the free Hamiltonian. For the theory (\ref{iphi3}), such $S$ matrix is not
unitary. Indeed, for a unitary theory, the imaginary part of all forward scattering amplitudes should be positive.
For the $2 \to 2$ amplitude, it is still the case at least perturbatively --- the analytic expression
for the 1--loop amplitude  depicted in Fig. 1a is the same as in the theory $\gamma \phi^3$ with real 
$\gamma$
that is unitary at perturbative level. But the 1--loop $3 \to 3$ amplitude depicted 
in Fig. 1b has an opposite 
sign compared to the same amplitude in the theory $\gamma \phi^3$. This violates unitarity.

\section{Non-anticommutative WZ model.}
The main subject of the present paper are so called non-anticommutative (NAC) supersymmetric theories.  
They were introduced first in Ref.\cite{Seiberg}. 
Seiberg took the standard Wess-Zumino model
 \be
\label{WZ}
{\cal L} \ =\ \int d^4\theta \, \bar \Phi \Phi + \left[ \int d^2\theta \left( \frac {m\Phi^2}2
+ \frac {g \Phi^3}3
\right) + {\rm c.c} \right] \nn \\
\equiv \ |\partial_\mu \phi|^2 + i \bar \psi \hat {\partial} \psi  - |W(\phi)|^2 
+ \left[ W'(\phi) \psi^2 + H.c. \right]
 \ee
with $W(\phi) = m\phi + g\phi^2$  and deformed it by introducing the nontrivial anticommutator
  \be
\label{Calbet}
 \{\theta^\alpha, \theta^\beta \} \ =\ C^{\alpha\beta}\ ,
 \ee
$  C^{\alpha\beta} =  C^{\beta\alpha}$, in the assumption that all other (anti)commutators vanish,
  \be
\label{drugiekom}
\{\bar \theta^{\dot \alpha}, \bar \theta^{\dot \beta} \} = \{ \theta^{\alpha}, \bar \theta^{\dot \beta} \} =
[\theta^\alpha, x^L_\mu] = [ \bar \theta^{\dot \alpha}, x^L_\mu] = [x^L_\mu, x^L_\nu] = 0\ .
  \ee
Note that this all was written in the {\it chiral} basis,
$x_\mu^L = x_\mu^{\rm central} + i\theta \sigma_\mu \bar \theta $.

The anticommutator
(\ref{Calbet}) introduces a constant self-dual tensor, which explicitly breaks Lorentz invariance.
However, the deformed Lagrangian expressed in terms of the component fields proves still
to be Lorentz invariant. Indeed, it is easy to find that the
 kinetic term $\int d^4\theta \, \bar \Phi \Phi $ is undeformed and the only extra piece comes from
 \be
 \label{F3}
\Delta {\cal L} =    \frac g 3 
\int d^2\theta \, \Phi*\Phi*\Phi - \frac g 3 \int d^2\theta \, \Phi^3 = 
 -   \frac g 3  \,\det \|C\| F^3\ ,
 \ee
$F$ being the auxiliary field.
 It depends only on
the scalar $~\det\|C\|$ and is obviously Lorentz invariant. Adding the usual terms 
$  F(m \phi + g \phi^2)  +  \bar F(m \bar \phi + \bar g \bar\phi^2) $ \ coming from superpotential 
 and $F\bar F$ from the kinetic term, and
expressing $F$ and $\bar F$ via
$\phi$ and $\bar\phi$, we see that the undeformed potential $|m\phi + g \phi^2|^2$
acquires an extra holomorphic contribution $\propto g(m\bar\phi + \bar g \bar\phi^2)^3$.

This extra contribution is {\it complex}. This makes the Lagrangian and Hamiltonian complex. A generic complex
Hamiltonian has  complex spectrum, the corresponding evolution operator is not unitary,  and
the theory has  little physical sense. This is the reason why people mainly considered up to now NAC models
 in Euclidean space only (where these problems do not arise) and were not interested in the corresponding Minkowski dynamics. 
We will show, however, that, even though the Hamiltonian of NAC WZ model does not look Hermitian, it is
in fact crypto-Hermitian under a special
choice of the deformation parameter and posseses a real spectrum.

This is exactly what was observed in Ref. \cite{IS} for another NAC deformed model, 
the Aldrovandy-Schaposnik model
\cite{AS} (basically, this is a deformation of Witten's SQM model). We conjectured in Ref. \cite{IS} that
crypto-Hermiticity of the deformed Hamiltonian holds also in the NAC WZ model. We confirm this conjecture here.
Another conjecture of Ref. \cite{IS} that, in the WZ case, the spectrum of the deformed model is not shifted 
compared to the undeformed case is correct only when  
  $g = 0$. Then  the undeformed model is free and so is the deformed one. 
But the interacting model {\it is} deformed  in a nontrivial way and its spectrum is shifted.

Our second observation is that  the deformed model still has {\it four} conserved supercharges 
$Q_\alpha, \bar Q^{\dot \beta}$.
Half of the supercharges (the supercharges $Q_\alpha$  under the standard convention)
 are the same as in the undeformed model, while the supercharges  $\bar Q^{\dot \beta}$ are modified and acquire
 a rather complicated form. 
 The spectrum 
of the theory includes 2 bosonic vacuum states and degenerate quartets of excited states,
 like in the undeformed WZ theory.  In other words, the common lore that NAC deformations break
 half of supersymmetries is not correct (at least, it is not correct in the cases analyzed).
 All supersymmetries stay intact. Half of them are  manifest and another half are realized in a complicated
indirect way. 

Again, this is very much similar to what happens in the AS model where one of the supercharges has the 
same form as in undeformed Witten's model while another one is modified.   
 
Generically,  $Q_\alpha$ and $\bar Q^{\dot \beta}$ are not adjoint to each other and the Hamiltonian 
is not Hermitian. However, when the deformation parameter is chosen in a special way, the  spectrum  
stays real and the Hamiltonian
is thus crypto-Hermitian. On the other hand, the deformed field interacting theory is not unitary 
in the same sense and by the same 
reasons as $i\phi^3$ theory discussed above.
 For the  free WZ model,  the deformed model is physically
equivalent to the underformed one and  its $S$-matrix is trivial.

 To derive all the results mentioned above in a manifest way, let us consider 
 the dimensionally reduced system and 
assume that the fields do not depend on 
spatial coordinates. The reduced  Hamiltonian is  
  \be
\label{Ham}
H \ =\ \bar \pi \pi +  \bar \phi  \phi +   g \phi^2 \bar \phi +  \bar g \bar \phi^2 \phi   + 
g\bar g \bar \phi^2 \phi^2  -(1+ 2g\phi) \psi_1 \psi_2 - (1 + 2 \bar g \bar \phi) \bar \psi_2 \bar \psi_1 
\nonumber \\
  + \beta ( \bar \phi + \bar g \bar \phi^2)^3 
 \ee
with $\bar \psi_\alpha \equiv \partial/\partial \psi_\alpha$ and $\beta = g \det\|C\|/3$ being the deformation
parameter. For simplicity, we have set $m=1$. 

 The wave functions for this Hamiltonian have four components, being  represented as
 \be
\label{Psi}
 \Psi(\phi, \bar\phi, \psi_\alpha) = A(\bar\phi, \phi)  + 
B_\alpha (\bar\phi, \phi)  \psi_\alpha + C (\bar\phi, \phi) \psi_1\psi_2\ .
 \ee
  In the undeformed case, the Hamiltonian (\ref{Ham}) admits conserved supercharges
 \be
\label{superQ}
Q_\alpha = \pi\psi_\alpha + i\epsilon_{\alpha\gamma} \bar \psi_\gamma 
( \bar \phi  +  \bar g \bar \phi^2 )\, ,  \nonumber \\
\bar Q_\beta = \bar \pi \bar \psi_\beta - i\epsilon_{\beta\delta} \psi_\delta (\phi +  g \phi^2 )
 \ee
with  $\epsilon_{12} = 1$. They satisfy the usual ${\cal N} = 2$ SQM algebra
 \be
\label{alg}
\{Q_\alpha, Q_\beta \} = \{\bar Q_\alpha, \bar Q_\beta \} = 0, \ \ \ \{Q_\alpha, \bar Q_\beta \} 
= H\delta_{\alpha\beta}
 \ee
  
Consider first the free Hamiltonian 
 \be
\label{Ham0}
H_0 \ =\ \bar \pi \pi +  \bar \phi  \phi   -   \left( \psi_1 \psi_2 + \bar \psi_2 \bar \psi_1 \right) \ .
 \ee
 This is the supersymmetric 2-dimensional oscillator and the wave functions can be
found explicitly being expressed via certain Laguerre polynomials. 
Let the functions $|ln \rangle$ be 
the eigenfunctions
of the bosonic Hamiltonian $  \bar \pi \pi +  \bar \phi  \phi  $  
($l$ is the eigenvalue of the charge operator 
$i(\phi\pi - \bar \phi \bar \pi) $ that commutes with $H_0$, 
  and $n$ is the principal quantum number). 
The explicit expressions for first few levels are
 \be
\label{levels}
|00\rangle \ =\ \sqrt{\frac 2\pi} e^{-\bar \phi \phi }, \ \ \ |10\rangle \ =\ \sqrt{\frac 2\pi} 
\phi e^{-\bar \phi \phi }, \ \ \ \ 
|\!-\!\!10\rangle \ =\ \sqrt{\frac 2\pi} \bar \phi e^{-\bar \phi \phi }, \nonumber \\
|01\rangle \ =\  \sqrt{\frac 2\pi} (1-2\phi \bar \phi)  e^{-\bar \phi \phi },\ \ \ \ |20\rangle \ =\ 
\frac 2{\sqrt{\pi}} \phi^2 e^{-\bar \phi \phi },
\ \ \ \ |\!-\!\!20\rangle \ =\ \frac 2{\sqrt{\pi}} \bar\phi^2 e^{-\bar \phi \phi }, \ldots
 \ee
The spectrum of the full  Hamiltonian (\ref{Ham0}) 
involves the fermionic states $\Psi^{\rm ferm} = |ln \rangle \psi_\alpha  $ with the energies 
$E^F_{ln} = |l| + 2n +1$ 
and the bosonic states 
 \be
\label{vac}
\Psi^{\rm bos} = |ln\pm \rangle \ \equiv \ \frac 1{\sqrt{2}} \left(1 \pm \psi_1 \psi_2) \right) |ln\rangle
 \ee  
with the energies $E_{nl+}^B = |l| + 2n,\ \ E_{nl-}^B = |l| + 2n + 2$. There is a single vacuum state 
$|00+\rangle$, while the excited
states come in quartets: the quartet of states 
\be
\label{E1}
|00\rangle \psi_\alpha, \ \ \ \ \ |\pm\!1 0+\rangle
 \ee
has the energy $1$, there are two quartets of energy $2$:
 \be
\label{E2}
\left\{|\!-\!\!20+\rangle, \  |\!-\!\!10\rangle \psi_\alpha,\  
\frac { \left[ |00-\rangle + |01+\rangle \right]}{\sqrt{2}} \right\}
 \ {\rm and} \ 
\left\{ \frac { \left[ |00-\rangle - |01+\rangle \right]} {\sqrt{2}} , \   |10\rangle \psi_\alpha, \   
|20+\rangle \right\} \ ,
 \ee
three quartets of energy $3$, etc.
The members of a quartet are produced from each other by the action of the supercharges (\ref{superQ})

Let us assume $g$ and $\beta$ to be nonzero but small and treat them perturbatively. To determine the  energy shifts 
 is a not so difficult exercise in standard quantum mechanics 
perturbation theory. One has to 
evaluate the graphs in Fig.2, 
 where  the charge -3 of the perturbation $\beta  \bar\phi^3$ 
(in the lowest nontrivial order, we can set $g=0$ in the last term in Eq.(\ref{Ham}))
is compensated by three insertions
of the perturbation $   g \bar \phi \phi^2 - 2 g  \phi  \psi_1  \psi_2$ of charge 1, 
while the perturbations $\propto \bar g$ and
$\propto \bar g g $  in Eq.(\ref{Ham})  are not relevant to this order.

\begin{figure}[h]
   \begin{center}
 \includegraphics[width=4.0in]{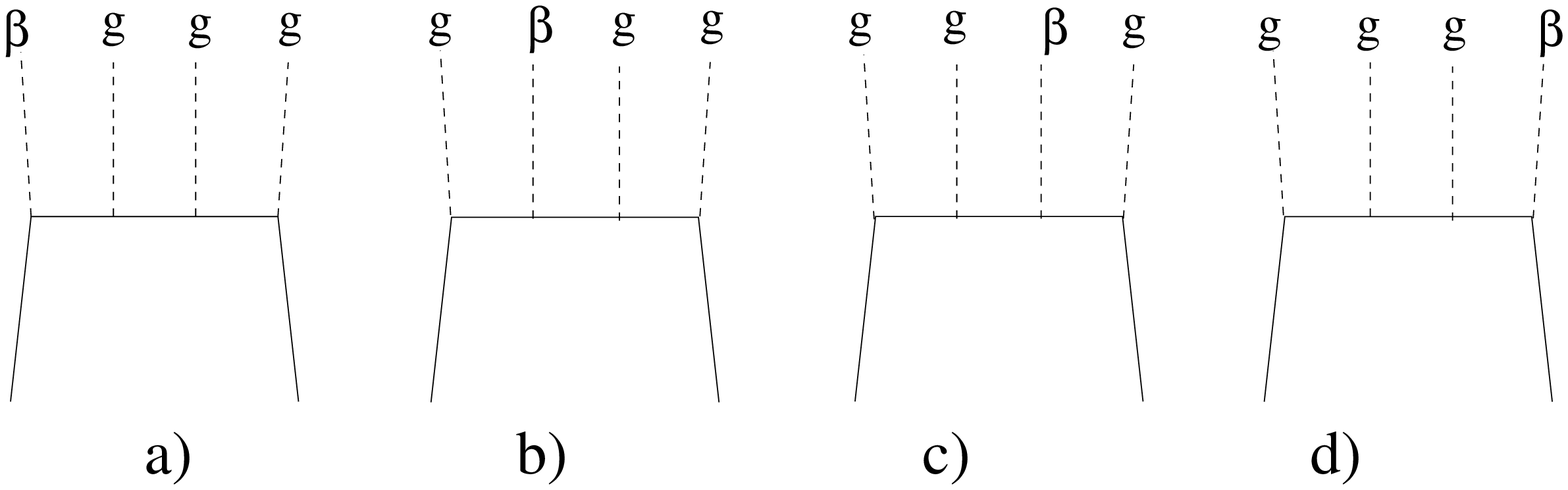}
    \end{center}
\caption{Graphs contributing to the energy shift $\propto \beta  g^3$}
\label{bggg}
\end{figure}

A somewhat long, but straightforward calculation gives the result: the ground state energy is not shifted and stays zero
\footnote{The Witten index \cite{Witten} of the interacting Wess-Zumino model is not 1 but 2, 
 and the second vacuum state should
appear. But the corresponding wave function lives at large values of $|\phi|$ and its presence cannot be detected 
perturbatively.},   
while the first excited level is shifted by
 \be
\label{DE1}
 \Delta E_1 = - \frac {155}{36} \beta  g^3   \ .
 \ee
The shift is the same for the bosonic states $|\!\pm\!10+\rangle$ and the fermionic states $|00\rangle \psi_\alpha$. 
Generically, the energy shift is complex, but, for real $\beta  g^3$, it is real.

The deformation $\propto W^3(\bar \phi)$ follows from NAC  machinery, but one can equally well consider 
simpler complex deformations $\Delta H = \beta_1 \bar\phi$ and $\Delta H = \beta_2 \bar\phi^2$. 
Again, the vacuum energy is not shifted, 
while the excited levels are shifted
such that the  quartets of the states are still degenerate. For the first excited quartet with energy $E_1 = 1$, one can calculate
(it is  the second order of perturbation theory for the deformation $\propto \beta_1$ and the third order for the deformation
$\propto \beta_2$)
 \be
\Delta E_1 (\beta_1) \ =\ -\beta_1  g; \ \ \ \ \ \ \ \ \   \Delta E_1 (\beta_2) = \frac {31}{18}   \beta_2  g^2 \ .
  \ee
The zero vacuum energy and 4-fold degeneracy  of the excited levels means that the deformed model still enjoys supersymmetry and 
the algebra (\ref{alg}) holds, though $H$ is not necessarily Hermitian and $Q_\alpha$ and $\bar Q_\alpha$ are not necessarily 
conjugate to each
other (cf. \cite{Mostsus}, \cite{AS}).  

Speaking of the supercharge $ Q_\alpha$, it is still given by the expression in 
Eq.(\ref{superQ}) , which commutes
with the  deformed Hamiltonian. On the other hand, the commutator of the undeformed supercharge $\bar Q_\alpha$ with the deformed 
Hamiltonian does not vanish. In contrast to AS model where a simple expression for the deformed supercharge
$\bar{Q}_\alpha^{\rm deformed}$ can be written \cite{AS}, we cannot do it in our case. By no means $\bar{Q}_\alpha$
can be obtained by complex conjugation 
of the supercharge  $ Q_\alpha$. Indeed, a pair of complex conjugate supercharges would mean Hermiticity of Hamiltonian, 
but the Hamiltonian (\ref{Ham}) is not  Hermitian. 
The fact that its spectrum is real (when $\beta g^3$ is real) tells, however, that the Hamiltonian is crypto-Hermitian
in the same sense as the AS Hamiltonian is. In particular, the operator $R$ rotating the Hamiltonian
to the manifestly Hermitian form should exist.

Even though explicit expressions for $\bar Q_\alpha$ are not known, one can argue 
that the quartet supersymmetric structure of the spectrum must hold without making explicit calculations. It 
can be reconstructed 
(at least, perturbatively \footnote{It would be very interesting to study the spectrum of the deformed 
Hamiltonian numerically. One cannot  
exclude a possibility that exceptional points \cite{Heiss} in the  space of couplings   appear such that 
the supersymmetric structure of the spectrum would be lost for large enough values of $\beta, g$.}) 
using only $ Q_\alpha$ and not $\bar Q_\alpha$. 
 Indeed, for each supersymmetric quartet of the eigenstates of the free Hamiltonian $H_0$,  
a member $\Psi$ annihilated by the action of $\bar Q_\alpha$, but not $ Q_\alpha$, can be chosen. 
 Three other members of the quartet are $ Q_{1,2}\Psi$ and
$ Q_1  Q_2 \Psi$. Let $\tilde \Psi$ be the corresponding eigenstate of the full Hamiltonian 
(when $\beta$ and $g$ are small, one can be sure
that such state exists). Then $\tilde \Psi$,  $ Q_\alpha \tilde \Psi$, and $ Q^2 \tilde\Psi$ 
represent a quartet of degenerate eigenstates of the interacting
deformed Hamiltonian. Once the states are known, the matrix elements of $\bar Q_\alpha$ can be defined 
to be equal to the corresponding matrix elements
in the free undeformed basis multiplied by $\sqrt{E_n^{\rm exact}/E_n^{\rm free}}$. 

If you will, the theories of this kind (where supersymmetry is not manifest at the Lagrangian level, but 
is there as far as the structure of the spectrum is concerned) can be called {\it cryptosupersymmetric}.
A very interesting question to be studied is whether and if so then how cryptosupersymmetry of deformed 
models (i.e. usual supersymmetry with complicated deformed supercharges) corresponds to 
the so called  {\it twist-deformed supersymmetry} for conventional supersymmetry generators unravelled in Refs. \cite{twist}.
(It was shown there that the conventional generators, which do not satisfy standard SUSY algebra in the deformed case, form
a certain twisted Hopf quantum superalgebra.) 

\section{Discussion.}

What conclusions concerning NAC field theories can be made on the basis of this analysis ?
If we put the theory in a finite spatial box and be interested in the spectrum of the Hamiltonian
thus obtained, one can conjecture that its properties should
 be similar to the properties of the dimensionally reduced deformed WZ Hamiltonian:
 \begin{itemize}
\item The ground state energy(ies) is(are) still zero (if supersymmetry is not spontaneously broken)
and the $2^{\cal N}$ degeneracy  of the excited spectrum states should be kept.
 \item For certain values of the deformation parameters and the couplings, the spectrum of the deformed Hamiltonian
 should enjoy crypto-Hermiticity property.  
\end{itemize}

 However,   a conventionally defined $S$-matrix is not unitary in deformed interacting NAC theories by the same token as it 
is not unitary in the theory $i\gamma \phi^3$. The complexity of Minkowski space Lagrangian
strikes back at this point. 
This means that these  theories cannot be attributed
a {\it conventional} physical meaning. More studies of this question are necessary. Maybe even if $S$-matrix of the
theory is not unitary, unitarity of its finite time finite box evolution operator (that follows from crypto-Hermiticity
of the Hamiltonian) suffices to  make the theory meaningful ? A positive answer to this question would mean a breakthrough
in understanding of not only  NAC theories, but also theories with higher derivatives in the Lagrangian. In Ref.\cite{TOE},
we argued that the fundamental Theory of Everything may be a theory of this kind. We address the reader to this paper and 
also to the papers \cite{brmog1} for discussions and speculations on this subject.

\end{document}